\documentclass{article}
\usepackage[ansinew]{inputenc} % Acepta caracteres en castellano
\usepackage{graphicx}
\usepackage{geometry}      % See geometry.pdf to learn the layout options.
\usepackage{amsmath}
\usepackage{epsfig}

\title{General Relativistic Radiant Shock Waves in the Post-Quasistatic Approximation}

\author{Jorge A Rueda H\footnote{jrueda@ula.ve} and L A N\'u\~nez\footnote{nunez@ula.ve}
\\\\$^{*}$ Centro de F\'isica Fundamental, Universidad de Los Andes, M\'erida 5101, Venezuela \\
Escuela de F\'isica, Universidad Industrial de Santander,A.A. 678,
Bucaramanga, Colombia\\\\$^{\dag}$ Centro de F\'isica Fundamental,
Universidad de Los Andes, M\'erida 5101, Venezuela \\ Centro
Nacional de C\'alculo Cient\'ifico, Universidad de Los Andes,
\textsc{CeCalCULA,}\\ Corporaci\'on Parque Tecnol\'ogico de
M\'erida, M\'erida 5101, Venezuela}

\date{November 2006}

\begin{document}

\maketitle

\begin{abstract}
An evolution of radiant shock wave front is considered in the
framework of a recently presented method to study self--gravitating
relativistic spheres, whose rationale becomes intelligible and finds
full justification within the context of a suitable definition of
the post--quasistatic approximation. The spherical matter
configuration is divided into two regions by the shock and each side
of the interface having a different equation of state and
anisotropic phase. In order to simulate dissipation effects due to
the transfer of photons and/or neutrinos within the matter
configuration, we introduce the flux factor, the variable Eddington
factor and a closure relation between them.  As we expected the
strength of the shock increases the speed of the fluid to
relativistic values and for some critical ones is larger than light
speed. In addition, we find that energy conditions are very sensible
to the anisotropy, specially the strong one. As a special feature of
the model , we find that the contribution of the matter and
radiation to the radial pressure are the same order of magnitude as
in the mant as in the core, moreover, in the core radiation pressure
is larger than matter pressure.
\end{abstract}

%%%%%%%%%%%%%%%%%%%%%%%%%%%%%%%%%%%%%%%%%%%
\section{Introduction}
%%%%%%%%%%%%%%%%%%%%%%%%%%%%%%%%%%%%%%%%%%%
During the implosion that forms a relativistic compact object,
nearly all of its gravitational binding energy ($ ({GM^{2}} )/{R}
\sim 5\times10^{53} $ ergs $\sim0.2Mc^{2}$) is stored as internal
energy of a proto-neutron star and driven by neutrino diffusion.
Roughly speaking, there is a consensus that this collapsing scenario
requires, three main ``ingredients''
\cite{BruennDeNiscoMezzacappa2001}:
\begin{enumerate}
  \item a copious emission of radiation, been a consequence of the microphysics of the system,  tends to abandon the system, but the absorption and the scattering in the medium hinder it to escape freely.
  \item phase transitions that can induce local anisotropic pressures  (i.e. $P_{r} \neq P_{\perp}$). An increasing amount of theoretical evidence strongly suggests that,
for certain density ranges, a variety of very interesting physical
phenomena may take place giving rise to local anisotropy (see
\cite{BowerLiang1974,HerreraSantos1997} and
\cite{HerreraMartinOspino2002,HerreraEtal2004,ChaisiMaharaj2005,ChaisiMaharaj2006,DevGleiser2002,DevGleiser2003,Ivanov2002,MakHarko2002,MakHarko2003,SharmaMukherjee2002}
for more recent studies ).
  \item the formation and propagation of a surface of discontinuity: a  shock wave, a detonation or combustion front (deflagration or slow combustion) having width very small compared to the size of the system.
\end{enumerate}
Although now exist several independent numerical codes which provide
accurate modeling of gravitational collapse in full General
Relativity (see \cite{Font2003} for a good review on this subject
and/or visit some these links concerning codes for simulations
\cite{GRAstro3D} ), none of these codes provide all the above
mentioned ``ingredient''.

In order to explore the influence that the scheme of radiation and
the local anisotropy exert on the propagation of a surface of
discontinuity, we shall follow a ``seminumerical'' approach which
starting from a known interior (analytical) static spherically
symmetric ( considered as ``seed'') solution to the
Tolman-Oppenheimer-Volkov equation.  It can be considered as a
continuation of previous studies \cite{HerreraNunez1987,
HerreraNunez1989,HerreraBarretoNunez1991}. This scheme transforms
the Einstein partial differential equations into a system of
ordinary differential equations for quantities evaluated at the
surfaces (boundary \& shock). It is an extension of the so called
HJR method \cite{HerreraJimenezRuggeri1980}, which has been
successfully applied to a variety of astrophysical scenarios (see
\cite{HerreraNunez1990} and references therein) and which has been
recently revisited
\cite{BarretoMartinezRodriguez2002,HerreraEtal2002,HerreraDiPriscoBarreto2006}.

When considering the interaction between radiation and ultradense
matter, we consider two quantities: the \textit{flux factor,} $f =
{\mathcal F}/{\rho_{R}},$ and the \textit{variable Eddington
factor,} $\chi= P_R/{\rho_{R}}, $ and a closure relation between
them, i.e., $\chi=\chi(f)$  (see
\cite{Dominguez1997,PonsIbanezMiralles2000,SmitVandenHornBludman2000,AguirreNunezSoldovieri2005}
), where ${\cal F}$, $\rho_R$ and $P_R$ are the radiation flux, and
the contribution of the radiation to the energy density and radial
pressure.

Here we shall show that the strength of the shock appears to be a
very significant feature concerning the evolution of the
distribution, because it drastically increases the fluid velocity
behind the shock.  We have also found that the energy conditions
(specially the strong one) are very sensitive to the variation of
the local anisotropy, and hydrodynamic \& radiation  pressures
emerge of the same order of magnitude and even, at the core,
radiation pressure is larger than that of matter counter part.

This paper is organized as follows: in section \ref{fieldequations}
the energy-momentum tensor and field equations for a non-static
spherically symmetric fluid with matter and radiation are
established. The Post-Quasistatic approximation is considered in
section \ref{PQA} Section \ref{shockwavemodel} is devoted to
describe the Taub junction conditions across the surface of
discontinuity.  Finally, we end this work with discussing some the
results of the numerical simulations.

%%%%%%%%%%%%%%%%%%%%%%%%%%%%%%%%%%%%%%%%%%%
\section{Matter, radiation and field equations}\label{fieldequations}
%%%%%%%%%%%%%%%%%%%%%%%%%%%%%%%%%%%%%%%%%%%

\begin{figure}
\epsfig{width=18pc,file=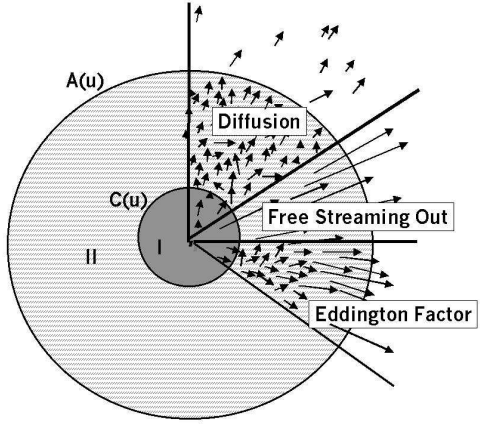} \hspace{2pc}
\begin{minipage}[b]{18pc}
\caption{Regions of the matter distribution. The core is labeled by
$I$ is the more compact one; the mantle ($II$) is the region in the
middle and the exterior space--time ($III$).} \label{esfeddf}
\end{minipage}
\end{figure}

As it can be appreciated in Figure \ref{esfeddf}, the collapsing
configuration is described by three regions: a core ($I$) as the
more inner region, the mantle ($II$) in the middle, and the outer
space labeled by ($III$).

Core and mantle are separated by mean of a shock (discontinuity
hypersurface) and a boundary surface (hypersurface with velocity
equal to the fluid in these point) is define the matter
configuration.  Both the shock and the boundary surface are
time-like hypersurfaces and there only exists unpolarized radiation
emerging which is modeled by the radiative Vaidya exterior metric.

The line elements at each region are
\begin{equation}
ds^2_{I,II}=e^{\nu} dt^2-e^{\lambda}dr^2-r^2(d\theta^2+\sin^2\theta
d\phi^2),
\end{equation} and
\begin{equation}
ds^2_{III}=\left(1-\frac{2 M(u)}{R}\right)du^2+2 du
dR-R^2(d\theta^2+\sin^2\theta d\phi^2)\, ,
\end{equation}
where $\nu$ and $\lambda$ are functions of $t$ and $r$, $u$ is the
retarded time and $R$ is a null coordinate ($g_{RR}=0$).

For a comoving observer with radial velocity $\omega$ relative to
local Minkowskian coordinates, the physical content of the fluid is
represented by the radiation flux ${\cal F}$ in the radial
direction; the energy density $\bar{\rho}$ and  $\bar{P}_r$, and
$\bar{P}_{\bot}$ as the radial and tangential pressures,
respectively.  The bar stands for the total (hydrodynamic $+$
radiation) contribution. Thus, the energy-momentum tensor for each
region can be written as,
\begin{eqnarray}
(T_{\alpha
\beta})_{I,II}&=&(\bar{\rho}+\bar{P}_\bot)u_{\alpha}u_{\beta}-\bar{P}_\bot
g_{\alpha \beta}+(\bar{P}-\bar{P}_\bot)\chi_{\alpha}\chi_{\beta}+2
{\cal F}_{(\alpha}u_{\beta)} \nonumber\\
(T_{\alpha \beta})_{III}&=&-\frac{1}{4 \pi R^2} \frac{dM}{du}
\delta^{0}_{\alpha} \delta^{0}_{\beta}
\end{eqnarray}
where \cite{MihalasMihalas1984}
\begin{eqnarray}
\bar{\rho}&=&\rho+\rho_R\, ,\quad \bar{P}=P_r+P_R\, ,\quad
\bar{P}_\bot =P_\bot+(P_\bot)_R\, ,\quad (P_\bot)_R = \frac{\rho_R-P_R}{2}\, ,\\
u_{\alpha}&=&\frac{e^{\nu/2}\delta^{0}_{\alpha}-\omega
e^{\lambda/2}\delta^{1}_{\alpha}}{\sqrt{1-\omega^2}}\, ,\quad
\chi_{\alpha}=\frac{-\omega e^{\nu/2}\delta^{0}_{\alpha}+
e^{\lambda/2}\delta^{1}_{\alpha}}{\sqrt{1-\omega^2}}\, ,\quad {\cal
F}_{\alpha}=-{\cal F}\chi_{\alpha}\, ,
\end{eqnarray}
with the subscript $R$ denoting the contribution of the radiation.

For the above energy-momentum tensor, Einstein field equations
$G^{\alpha}_{\hphantom{\alpha}\beta}=-8\pi
T^{\alpha}_{\hphantom{\alpha}\beta}$ become
\begin{equation}\label{field1}
-8\pi \left( \frac{\bar{\rho}+ \omega^2 \bar{P} +2 \omega {\cal
F}}{1-\omega^2} \right)=-\frac{1}{r^2}+e^{-\lambda}\left(
\frac{1}{r^2} - \frac{\lambda'}{r} \right)\, ,
\end{equation}
\begin{equation}\label{field2}
-8\pi \left( \frac{\bar{P}+\omega^2 \bar{\rho} + 2 \omega {\cal
F}}{1-\omega^2} \right)=-\frac{1}{r^2}+e^{-\lambda}\left(
\frac{1}{r^2} + \frac{\nu'}{r} \right)\, ,
\end{equation}
\begin{equation}\label{field3}
-8\pi \bar{P}_{\bot}=\frac{e^{-\nu}}{4}\left[ 2\ddot{\lambda} +
\dot{\lambda}(\dot{\lambda}-\dot{\nu}) \right] -
\frac{e^{-\lambda}}{4}\left[ 2\nu'' +
\left(\nu'+\frac{2}{r}\right)(\nu'-\lambda') \right] \, ,
\end{equation}
\begin{equation}\label{field4}
-8\pi e^{\frac{\nu+\lambda}{2}}\left[\frac{
\omega(\bar{\rho}+\bar{P})+(1+\omega^2){\cal F}}{1-\omega^2}
\right]=\frac{\dot{\lambda}}{r}\, ,
\end{equation}
where dots and primes are denoting time and radial derivatives,
respectively.

%%%%%%%%%%%%%%%%%%%%%%%%%%%%%%%%%%%%%%%%%%%%%%%%%%%%%%%%%%%%%%%%%%%%%%%%%%%%%%%%%%%%
%%%%%%%%%%%%%%%%%%%%%%%%%%%%%%%%%%%%%%%%%%%%%%%%%%%%%%%%%%%%%%%%%%%%%%%%%%%%%%%%%%%%
\section{The Post--Quasistatic Approximation (PQA)}
\label{PQA}
%%%%%%%%%%%%%%%%%%%%%%%%%%%%%%%%%%%%%%%%%%%%%%%%%%%%%%%%%%%%%%%%%%%%%%%%%%%%%%%%%%%%
%%%%%%%%%%%%%%%%%%%%%%%%%%%%%%%%%%%%%%%%%%%%%%%%%%%%%%%%%%%%%%%%%%%%%%%%%%%%%%%%%%%%

The quasistatic regime is ``the next step coming out from
hydrostatic equilibrium''. In this regime the distribution changes
slowly in a typical time scale which is very long compared with the
characteristic scale within the sphere reacts to a perturbation.
Mathematically, it can be stated as \cite{HerreraEtal2002} \( {\cal
O}(\omega^2)=\ddot{\lambda}=\ddot{\nu}=\dot{\lambda}\dot{\nu}=\dot{\lambda}^2=\dot{\nu}^2=0\,
.\)

Now, we define the ``efective variables''
\begin{equation}
\tilde{\rho}=\frac{\bar{\rho}+\bar{P} \omega^2 +2 \omega {\cal
F}}{1-\omega^2} \qquad \mathrm{and}  \qquad
\tilde{P}=\frac{\bar{P}+\bar{\rho} \omega^2 +2 \omega {\cal
F}}{1-\omega^2}\, ,
\end{equation}
which satisfy the same set of equations as their corresponding
\textit{physical }variables in the quasistatic case.  Thus,  Herrera
\textit{et al.} \cite{HerreraEtal2002} define the post-quasistatic
regime as that corresponding to a system out of equilibrium (or
quasiequilibrium) but whose effective variables share the same
radial dependence as the physical variables in the state of
equilibrium (or quasi-equilibrium), i.e., is the closest possible
situation to a quasistatic evolution.

In this effective variables, the field equations
(\ref{field1})--(\ref{field4}) can be written as
\begin{eqnarray}
m &=& \int 4 \pi r^2 \tilde{\rho} dr\, ,\label{e1}\\
\nu &=& \int \frac{2(4 \pi r^3 \tilde{P} + m)}{r(r-2m)} dr\, ,\label{e2}\\
-8\pi \bar{P}_{\bot} &=& \frac{e^{-\nu}}{4}\left[ 2\ddot{\lambda} +
\dot{\lambda}(\dot{\lambda}-\dot{\nu}) \right] -
\frac{e^{-\lambda}}{4}\left[ 2\nu'' +
\left(\nu'+\frac{2}{r}\right)(\nu'-\lambda') \right] \, ,\label{e3}\\
\dot{m} &=& -\frac{4 \pi r^2 e^{\frac{\nu-\lambda}{2}}}{1+ \omega^2}
[\omega(\tilde{\rho} + \tilde{P}) + (1-\omega^2) {\cal F}]\,
,\label{e4}
\end{eqnarray}
where the mass function is defined by $ e^{-\lambda}=1-\frac{2
m(t,r)}{r}\, . $

%%%%%%%%%%%%%%%%%%%%%%%%%%%%%%%%%%%%%%%%%%%%%%%%%%
%%%%%%%%%%%%%%%%%%%%%%%%%%%%%%%%%%%%%%%%%%%%%%%%%%
\section{Collapsing Model with Shock wave in the PQA}
\label{shockwavemodel}
%%%%%%%%%%%%%%%%%%%%%%%%%%%%%%%%%%%%%%%%%%%%%%%%%%
%%%%%%%%%%%%%%%%%%%%%%%%%%%%%%%%%%%%%%%%%%%%%%%%%%

\subsection{The regions and the equations of state}
Following Herrera and collaborators
\cite{HerreraNunez1987,HerreraBarretoNunez1991}, we consider the
core equation of state inspired by the anisotropic
Schwarzschild-like interior solution, i.e.
\cite{BowerLiang1974,CosenzaEtal1980}
\begin{equation}
\tilde{\rho}_{I} = f(t) \qquad \mathrm{and}  \qquad
 \tilde{P}_{I} = \tilde{\rho}_{I}\left\{ \frac{3 (1-8/3 \pi
r^2\tilde{\rho}_{I})^{\xi_{I /2}} k(t) -1}{ 3-(1-8/3 \pi
r^2\tilde{\rho}_{I})^{\xi_{I /2}} k(t)} \right\}\, ,\label{P1}
\end{equation}
where $\xi_I=1-2h_I$ is the anisotropic parameter with $h_I=0$
corresponding to an isotropic model. For the mantle we have an
anisotropic Tolman VI-like solution, i.e.
\begin{equation}
\tilde{\rho}_{II} = \frac{3 g(t)}{r^2} \qquad \mathrm{and}  \qquad
\tilde{P}_{II} = \frac{\tilde{\rho}_{II}}{3} \left[
\frac{1-9D(t)r^{\sqrt{4-3\xi_{II}}}}{1-D(t)r^{\sqrt{4-3\xi_{II}}}}
\right]\, .\label{P2}
\end{equation}
again, $h_{II}$ (or $\xi_{II}=1-2h_{II}$) is the parameter messuring
the anisotropy of this region and the time-dependent functions
$f(t), k(t), g(t)$ and $D(t)$, are obtained from boundary
conditions.

\subsection{Junction Conditions}
The matching across the boundary surface to the Vaidya metric is
obtained though the Darmois-Lichnerowicz junction conditions
\cite{BonnorVickers1981}, while across the shock front via
relativistic Rankine-Hugoniot conditions \cite{Taub1948}.
 As it was pointed out by Bonnor and Vickers some years ago \cite{BonnorVickers1981},
 Darmois-Lichnerowicz conditions are not equivalent to the O'Brien and Synge ones.
 In the present case, O'Brien and Synge  conditions lead to a contact
discontinuity (not a shock) forcing  the fluid velocity to be
continuos across the surface $r=c(t)$. Thus denoting the boundary
surface by $\Sigma=r-a(t)=0$, the matching conditions lead
\begin{equation}
e^{\nu_{a(t)}}=e^{-\lambda_{a(t)}}=1-\frac{2 M(u)}{a(u)} \qquad
\mathrm{and}  \qquad \bar{P}_{a(t)}={\cal F}_{a(t)}.
\label{acoplesuperficie1}
\end{equation}
Now, if the shock front is described by $\Gamma=r-c(t)=0$, the
continuity of the second fundamental form  implies the continuity of
the mass flux across the shock, and the continuity of the first
fundamental form and the Rankine-Hugoniot conditions, lead to
\begin{equation}
[\nu]_{c(t)}=[\lambda]_{c(t)}= 0 \, , \qquad \left[
e^{(\nu-\lambda)/2} \left\{ \frac{
\omega(\bar{\rho}+\bar{P})+(1+\omega^2){\cal F}}{1-\omega^2}
\right\} -\dot{c} \tilde{\rho} \right]_{c(t)}=0
\label{acoplechoque2}
\end{equation}
\begin{equation}\label{acoplechoque3}
\mathrm{and}  \qquad \left[ \dot{c} e^{(\lambda-\nu)/2}\left\{
\frac{\omega(\bar{\rho}+\bar{P})+(1+\omega^2){\cal F}}{1-\omega^2}
\right\} -\tilde{P} \right]_{c(t)}=0\, ,
\end{equation}
with $[X]_{c(t)} \equiv X|_{r=c(t)+}-X|_{r=c(t)-}$, where $+$ stands
for ahead from the front shock and $-$ for behind. Now, by using the
field equation (\ref{e4}) in equations (\ref{acoplechoque2}) and
(\ref{acoplechoque3}) we obtain
\begin{equation}
[\dot{m}+4\pi r^2 \dot{c} \tilde{\rho}]_{c(t)}=0 \qquad \mathrm{and}
\qquad [\dot{m}\dot{c} e^{\lambda-\nu}+4\pi r^2
\tilde{P}]_{c(t)}=0\, . \label{RH2}
\end{equation}
Expanding the mass function in a Taylor series about the shock
front, we can get a relation between time and radial derivatives and
it implies the jump of the time-derivative of the mass across the
shock. i.e.
\begin{equation}\label{taylorfinal}
\dot{m}(t,r)|_{c(t)}=\dot{m}(t,c)-\dot{c}m'(t,r)|_{c(t)} \quad
\Longrightarrow [\dot{m}]_{c(t)}=[-4\pi r^2 \dot{c}
\tilde{\rho}]_{c(t)}\, ,
\end{equation}
and by using (\ref{RH2}) we obtain
\begin{equation}
\dot{c}=e^{\frac{\nu_{c(t)}-\lambda_{c(t)}}{2}}\sqrt{\frac{[\tilde{P}]_{c(t)}}{[\tilde{\rho}]_{c(t)}}}\,
. \label{shockspeed}
\end{equation}

%%%%%%%%%%%%%%%%%%%%%%%%%%%%%%%%%%%%%%%%%%%%%%%%%%%%%%%%%%%%%%%%%%%%%%%%%%%%%%%%%%%%%%
%%%%%%%%%%%%%%%%%% Hasta ac'a creo que todo bien %%%%%%%%%%%%%%%%%%%%%%%%%%%%%%%%%%%%%

\subsection{Metric Functions and Effective Variables}

Let us to introduce the dimensionless variables
\begin{equation}\label{reescalamiento}
M = \frac{m_a}{m_a(0)}\, ,\quad A = \frac{a}{m_a(0)}\, ,\quad F =1 -
\frac{2 M}{A}\, ,\quad \Omega = \omega_a\, , \quad \mathrm{and}
\quad t \to \frac{t}{m_a(0)}\, ,
\end{equation}
where $m_a(0)$ is the total initial mass of the distribution, and
the physical variables
\begin{equation}\label{variablesfisicas}
\hat{E} = 4 \pi a^2 {\cal F}_{a}\, ,\quad L = -\dot{M}\, , \quad
\mathrm{and} \quad
 E= -\frac{\dot{M}}{F}\, ,
\end{equation}
with $\hat{E}$ the luminosity for a (non)comoving observer and $L$
the luminosity at infinity.

All physical variables can be obtained as functions of the above
surface variables $A, F, $ and $\Omega$. Thus, from the boundary
conditions (\ref{acoplesuperficie1}) we get
\begin{equation}\label{mye}
m_{II} = \frac{M r}{A}\, ,\qquad E = (1+\Omega)\hat{E}\, ,
\end{equation}
\begin{equation}\label{rho2def}
\tilde{\rho}_{II} = \frac{1-F}{8 \pi r^2}\, , \qquad \tilde{P}_{II}
= \frac{1-F}{24 \pi r^2}\left[ \frac{\psi - 9 \chi (r/a)^{\sqrt{4 -
3 \xi_{II}}}}{\psi - \chi (r/a)^{\sqrt{4 - 3 \xi_{II}}}} \right]\, ,
\end{equation}
\begin{equation}
e^{-\lambda_{II}} = F\, ,\qquad e^{\nu_{II}} = F \left\{ \frac{r}{a}
\left[ \frac{\psi - \chi(r/a)^{\sqrt{4-3\xi_{II}}}}{\psi -
\chi}\right]^{2/{\sqrt{4-3\xi_{II}}}} \right\}^{\frac{4(1-F)}{3F}}\,
,
\end{equation}
where \begin{equation*} \psi = 3(3+\Omega)(1-F) 6(1+\Omega)E\,
,\qquad \chi = (1+3\Omega)(1-F) - 6(1+\Omega)E\, .
\end{equation*}
Now, by introducing the shock ``force'' parameter $N$
$(\tilde{P}_I)_{c}= N (\tilde{P}_{II})_{c}$, we obtain
\begin{equation}\label{m1def}
m_{I} = \frac{M c}{A}(r/c)^{3}\, ,\qquad \tilde{\rho}_{I} =
\frac{3M}{4 \pi c^2 A}\, ,\qquad \tilde{P}_{I} =
\tilde{\rho}_{I}\left\{ \frac{3 u^{\xi_{I /2}} k(t) -1}{ 3-u^{\xi_{I
/2}} k(t)} \right\}\, ,
\end{equation}
\begin{equation}
e^{-\lambda_{I}} = 1 - \frac{2Mc}{A r} (r/c)^3\, ,\qquad e^{\nu_I}=H
u^{\Phi} (3-k u ^{\xi_I/2})^{8/\xi_I}\, ,
\end{equation}
where
\begin{equation*}
k=\frac{24 \pi c^2 \tilde{\rho}_I \beta+3 N \alpha
(1-F)}{F^{\xi_I/2}[72 \pi c^2 \tilde{\rho}_I \beta+ N \alpha
(1-F)]}\, ,\quad H=\frac{F^{1-\Phi}\left\{ \frac{c}{A} \left[
\frac{\beta}{8(F-1)} \right]^{2/\sqrt{4-3 \xi_{II}}}
\right\}^{\frac{4(1-F)}{3 F}}}{(3-k F^{\xi_I/2})^{8/\xi_I}}\, ,
\end{equation*}
\begin{equation*}
u=1-\frac{8\pi r^2\tilde{\rho}_{I}}{3}\, ,\quad
\Phi=\frac{1}{2}-\frac{3(1-F)}{16 \pi c^2 \tilde{\rho}_I}\, ,\quad
\alpha=\psi-9 \chi (c/A)^{\sqrt{4-3 \xi_{II}}}\, ,\quad
\beta=\psi-\chi (c/A)^{\sqrt{4-3 \xi_{II}}} \, .
\end{equation*}

Again, all effective variables and metric functions depend on $t$
through the surface variables $A, F$, $\Omega$ and $L$.

%\begin{figure}[h]
%\begin{minipage}{18pc}
%\includegraphics[width=18pc]{presionesenc.eps}
%\caption{ Hydrodynamic \& radiation pressures at both sides of the shock front. Notice that both pressures have the same order of magnitude and, at the core radiation pressure is larger.}
%\label{presionesenc}
%\end{minipage}\hspace{2pc}%
%\begin{minipage}{18pc}
%\includegraphics[width=18pc]{fomega.eps}
%\caption{Fluid velocity at $r = a(t)$. Notice that the boundary surface bounces and expands.}
%\label{fomega}
%\end{minipage}
%\end{figure}

\subsection{The Surface Equations}

In order to find the evolution of the surface variables, we have to
integrate a system of ordinary differential equations on $A, F,
\Omega$ and $L$, i.e.  \textit{the surface equations}. The first
surface equation can be obtained from the relation between the
coordinate velocity and the velocity of the comoving observer \(
\dot{r}= e^{\frac{\nu-\lambda}{2}} \omega\, , \) evaluated on
$\Sigma$.  The second equation emerges from time derivative of $F$
and by using the definition of the luminosity $L$
\begin{equation}\label{surface12}
\dot{A}=F \Omega\, , \qquad \dot{F}=\frac{(1-F)F\Omega+ 2 L}{A}\, ,
\end{equation}
where we have used the junction condition on $\lambda$ and $\nu$
(\ref{acoplesuperficie1}) and the dimensionless variables
(\ref{reescalamiento}).

The differential equation for $\Omega$ is obtained evaluating the
conservation law $T^{\alpha}_{r;\alpha}=0$, on $\Sigma$, i.e
\begin{equation}\label{TOV}
\tilde{P}' + \frac{(\tilde{\rho} + \tilde{P})(4 \pi r^3 \tilde{P} +
m)}{r(r-2m)} = \frac{2(\bar{P}_{\bot}-\tilde{P})}{r} +
\frac{e^{-\nu}}{4 \pi r (r-2m)}\left( \ddot{m} +
\frac{3\dot{m}^2}{r-2m} - \frac{\dot{m} \dot{\nu}}{2} \right)\, ,
\end{equation}
but we do not write it for its cumbersome form.

Because this system is overdeterminate, we should introduce one of
the ``surface'' functions. Since the only observable quantity
entering a ``real'' gravitational collapse is the luminosity, it
seems reasonable to provide such a profile as an input for our
modeling. Therefore, we select the luminosity profile to be a
Gaussian pulse centered at $t=t_0$
\begin{equation}
-\dot{M}=L=\frac{\Delta M_{rad}}{s \sqrt{2 \pi}}\exp\left[-
\frac{1}{2}\left( \frac{t-t_0}{s} \right)^2 \right]\, ,
\end{equation}
where $s$ is the width of the pulse and $\Delta M_{rad}$ is the
total mass lost in the process.

%We integrate the surface equations  for
%\begin{equation}
%A(0)=10.67 \quad \Omega(0)=-0.008 \quad c(0)=2.6
%\label{CondIni}
%\end{equation} with
%\begin{equation}
%N=2.8 \quad \xi_I=1.0 \quad \xi_{II}=0.8 \quad f_I=0.001 \quad f_{II}=1.0
%\label{Parametros}
%\end{equation} and
%\begin{equation}
%M(0)=2.04 \times 10^{5}M_{\bigodot}, \quad \Delta M_{rad}=0.01M(0), \quad t_0=15 \; s \quad \mathrm{and} \quad s=0.25 \; s
%\label{ParaPulso}
%\end{equation}

\subsection{The radiation hydrodynamic environment}

Conscious of the difficulties to cope with dissipation due to the
emission of photons and/or neutrinos, and aware of the uncertainties
of the microphysics when considering the interaction between
radiation and ultradense matter, we introduce a relation between the
radiation energy flux density and the radiation energy density, i.e.
the flux factor, $f={\cal F}/\rho_R$ , and the so called variable
Eddington factor, $\chi=P_R/\rho_R$, relating the radiation pressure
and the radiation energy density.

There are several of those closure relations reported in the
literature (see two recent comprehensive discussions on this subject
in \cite{PonsIbanezMiralles2000,SmitVandenHornBludman2000}). Few of
them are simply ad hoc relations that smoothly interpolate the
radiation field between the diffusive and free-streaming regimes.
Others, are derived from a maximum entropy principle
\cite{Dominguez1997} or assuming, angular dependence of the
radiative distribution functions. Even one of them has been
motivated from direct transport calculations.

For our simulations we shall use the Lorentz--Eddington closure
relation
\begin{equation}
\chi(f) = \frac{5}{3}-\frac{2}{3}\sqrt{4-3f^2}\, .
\end{equation}
Thus, we can to simulate any radiation phase between diffusion limit
and streaming out limit. Knowing radiation variables, the field
equations (\ref{field1})--(\ref{field4}) together with the
anisotropic equation of state, the physical variables $\rho$, $P_r$,
$P_{\bot}$, $\omega$, and ${\cal F}$ can be determined

\section{Analysis, and some prelimary results}

We integrate the surface equations  for
\begin{equation}
A(0)=10.67 \quad \Omega(0)=-0.008 \quad c(0)=2.6 \label{CondIni}
\end{equation} with
\begin{equation}
N=2.8 \quad \xi_I=1.0 \quad \xi_{II}=0.8 \quad f_I=0.001 \quad
f_{II}=1.0 \label{Parametros}
\end{equation} and
\begin{equation}
M(0)=2.04 \times 10^{5}M_{\bigodot}, \quad \Delta M_{rad}=0.01M(0),
\quad t_0=15 \; s \quad \mathrm{and} \quad s=0.25 \; s
\label{ParaPulso}
\end{equation}

\begin{figure}[h]
\begin{minipage}{18pc}
\epsfig{width=18pc,file=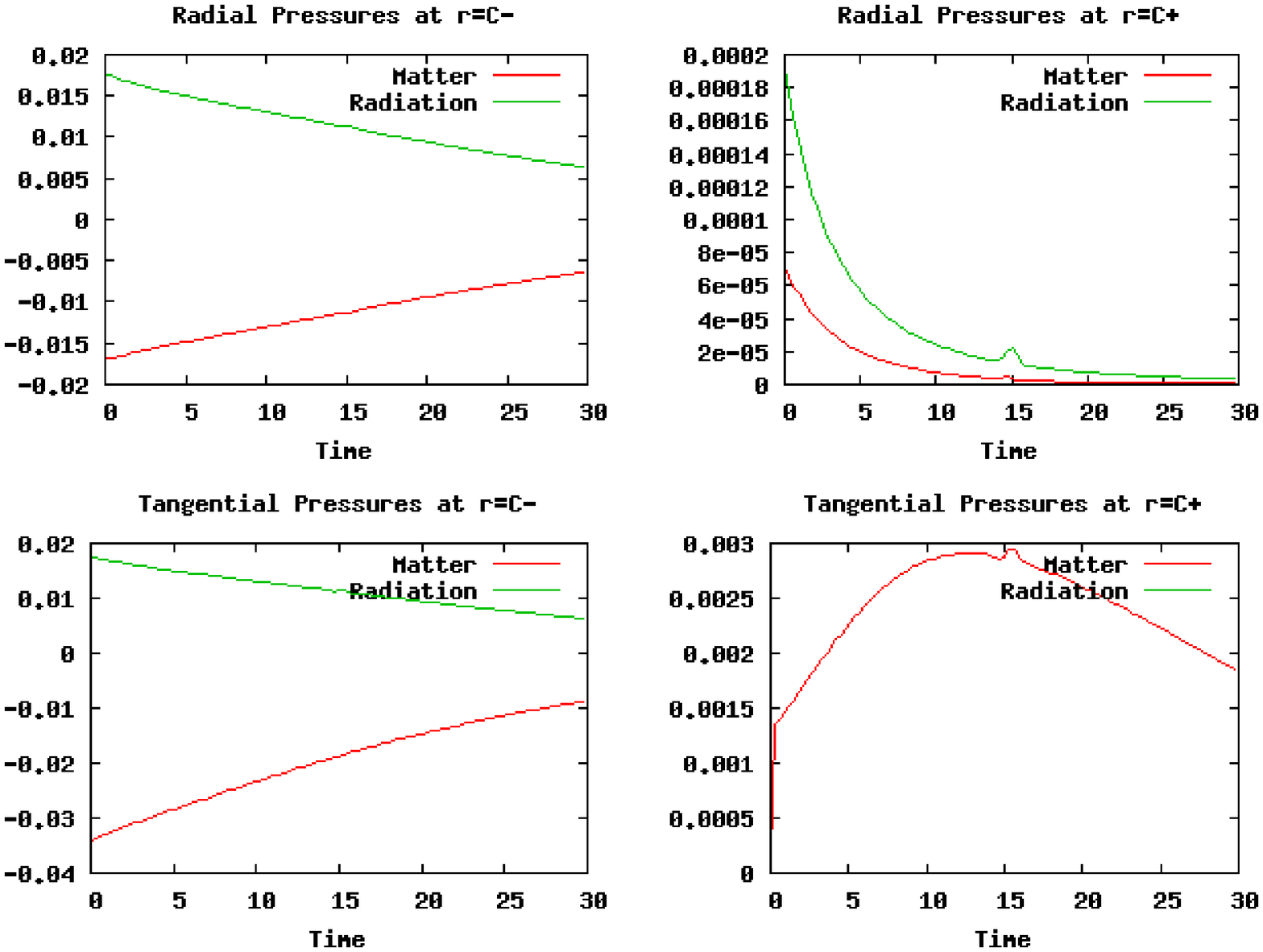} \caption{ Hydrodynamic \&
radiation pressures at both sides of the shock front. Notice that
both pressures have the same order of magnitude and, at the core
radiation pressure is larger.} \label{presionesenc}
\end{minipage}\hspace{2pc}%
\begin{minipage}{18pc}
\epsfig{width=18pc,file=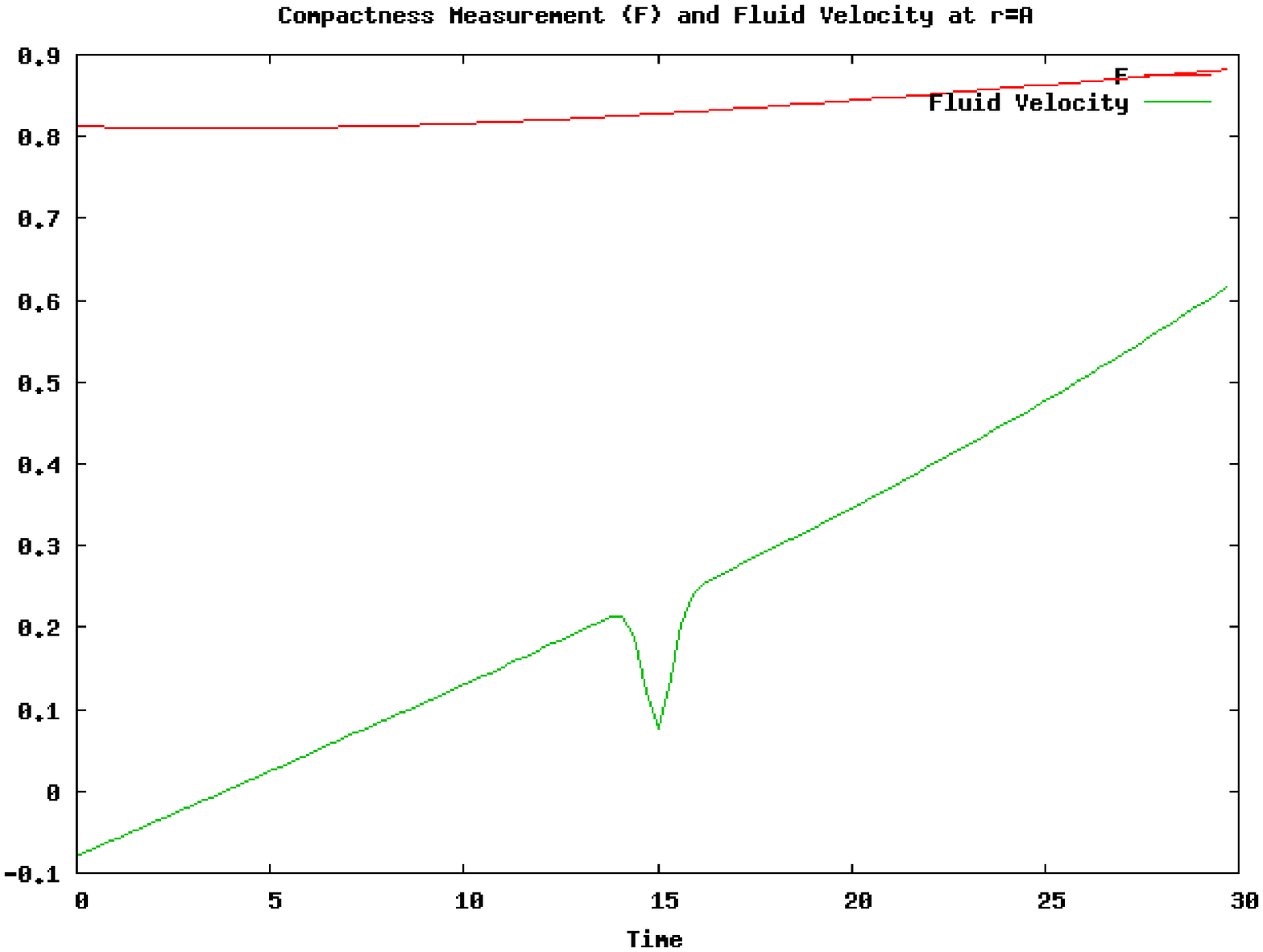} \caption{Fluid velocity at $r =
a(t)$. Notice that the boundary surface bounces and expands.}
\label{fomega}
\end{minipage}
\end{figure}

Our simulations show the energy conditions (specially the strong
one) are very sensitive to the variation of the local anisotropy.
We have found that the possible range for the anisotropic parameter
is  $1 \leq \xi \leq 4/3 $, with $\xi=1$ representing the isotropic
limit. Because the energy conditions, the behavior of the
distribution is different for the same absolute value of the
difference of anisotropy between the mantle and core
$|\xi_{I}-\xi_{II}|$ depending on which region is more anisotropic.
This was found for the case  $\xi_I=0.78$ and $\xi_{II}=0.98$ but
not for the inverse model ( $\xi_I=0.98$ and $\xi_{II}=0.78$). In
addition, for the same reason a very anisotropic core is not able to
maintain a very isotropic mantle.

Another special feature is that anisotropic cores force radiation
flux to decay slower than in the isotropic case and the velocity of
the fluid is slower than of the corresponding isotropic case.  It
seams to indicate that anisotropy slows down the condensation at the
core, maintaining it in non-ultrarelativistic regimes. This
situation has to be further explored in forthcoming studies.

As it can be appreciated from figure \ref{presionesenc}, it is found
that the hydrodynamic \& radiation  pressures become of same order
of magnitude and at the core, radiation pressure is larger. This is
consistent with the picture we expect for the radiation
hydrodynamics within a general relativistic gravitational collapse
\cite{MihalasMihalas1984}. Figure (\ref{fomega}) shows how the shock
front bounces, from a collapsing core it turns to expand violently.

The shock strength, $N$, also appears to be a very significant
feature concerning the evolution of the distribution, because it
drastically increases the fluid velocity behind the shock. From
equation (\ref{shockspeed}) we undertand that this parameter cannot
to be less than one because it leads to imaginary speed of the
shock.  The reason is that the fraction When we raise the shock
force parameter, the fluid velocity increases faster to relativistic
values and even for certain critical value (it depends on the whole
set of initial conditions) it reaches light speed rapidly.

As we have stressed, these are preliminary results that should
deserve further simulations.

\section*{Acknowledgments}
We gratefully acknowledge the financial support of the Consejo de
Desarrollo Científico Humanístico y Tecnológico de la Universidad de
Los Andes (CDCHT-ULA) under project C-1009-00-05-A, and to the Fondo
Nacional de Investigaciones Científicas y Tecnológicas (FONACIT)
under projects S1-2000000820 and F-2002000426.  J. A. R. H. thanks
to Universidad Industrial de Santander (Bucaramanga-Colombia) and
Universidad de Los Andes (Mérida--Venezuela) for the financial
support. \vspace{2pc}
%%%%%%%%%%%%%%%%%%%%%%%%%%%%%%%%%%%%%%%%%%%%%%%%%%%%%%%%%%%%%%%%%%%%%%%%%%%%% Aqui van las referencias %%%%%%%%%%%%%%%%%%%

%%%%%%%%%%%%%%%%%%%%%%%%%%%
% Las figuras
%%%%%%%%%%%%%%%%%%%%%%%%%%%
%\begin{center}
%\begin{tabular}{c}
%\includegraphics[width=10cm,height=10cm]{supmaschoque}\\
%\includegraphics[width=10cm,height=10cm]{velflujo}
%\end{tabular}
%\end{center}

%\begin{center}
%\begin{tabular}{c}
%\includegraphics[width=10cm,height=10cm]{densidades}\\
%\includegraphics[width=10cm,height=10cm]{presionesenc}
%\end{tabular}
%\end{center}

%\begin{center}
%\begin{tabular}{c}
%\includegraphics[width=10cm,height=10cm]{ptotalesenc}
%\end{tabular}
%\end{center}

%%%%%%%%%%%%%%%%%%%%%%%%%%%%%%%%%%

\end{document}